\documentclass[prd,aps,amsmath,amssymb,letter,showpacs]{revtex4}
\usepackage{graphicx}
\usepackage{epsfig,rotate,latexsym}
\begin{document}
\title{Enhancement of flow anisotropies due to  magnetic field 
in relativistic heavy-ion collisions}
\author{Ranjita K. Mohapatra}
\author{P. S. Saumia}
\author{Ajit M. Srivastava}
%
\affiliation{Institute of Physics, Sachivalaya Marg, 
Bhubaneswar 751005, India}
%
%

\begin{abstract}
It is known that the presence of background magnetic field in cosmic 
plasma distorts the acoustic peaks in CMBR. This primarily results from
different types of waves in the plasma with velocities depending
on the angle between the magnetic field and the wave vector. We 
consider the consequences of these effects in relativistic heavy-ion
collisions where very strong magnetic fields arise during early
stages of the plasma evolution. We show that flow coefficients can be 
significantly affected by these effects when the magnetic field remains
strong during early stages due to strong induced fields in the conducting 
plasma. In particular, the presence of magnetic field can lead to 
enhancement in the elliptic flow coefficient $v_2$.
\end{abstract}
\maketitle

 It has been recently shown by us that a deep correspondence exists
between the physics of inflationary density fluctuations in the early 
universe which result in the CMBR acoustic peaks and the physics of
flow in relativistic heavy-ion collision experiments (RHICE) \cite{cmbhic}. 
We further showed that important features such as acoustic peaks and
suppression of superhorizon fluctuations may be present in a plot of 
root-mean square values of flow coefficients $v_n^{rms}$.
The possibility of presence of this suppression of superhorizon 
fluctuations  has been recently pointed out by Sorensen in RHIC
data \cite{srnsn}. There have been several works now which discuss
flow fluctuations for a large range of values of $n$ \cite{flowfluct}. 
This non-trivial connection between the superhorizon fluctuations of 
inflationary universe and similar fluctuations in RHICE is now 
discussed in several works and its various consequences are explored 
(see, e.g. \cite{shuryak}). We mention 
that such a connection between physics of RHICE and that of inflationary
universe was never anticipated earlier, and indeed, at first sight, 
it looks surprising that a concept like superhorizon fluctuation which 
arises from highly non-trivial, superluminal expansion phase of the very 
early universe could have any relevance for relativistic heavy-ion 
collision experiments in laboratory.  Such superhorizon fluctuations in 
RHICE originate from the fact that in the center of mass frame the 
thermalization (and any local homogenization) happens rather quickly, 
within about 1 fm. Initial parton energy density distribution from HIJING  
shows that transverse fluctuations (arising from localization 
of partons inside initial nucleons, and from the fluctuations in 
nucleon coordinates) with wavelengths significantly larger than 1 fm are 
necessarily present at the time 1 fm even in central collisions. 

It was also emphasized in \cite{cmbhic} that various analysis tools of 
CMBR anisotropies can be effectively utilized for RHICE.  It was proposed 
in \cite{cmbhic} that instead of focusing on the average values of the flow 
coefficients $v_n$ for small values of $n$ \cite{v2fluct}, one should 
calculate root-mean square values of the flow coefficients $v_n^{rms}$ for a 
large range of $n$ upto 30-40. Further, these
calculations should be performed in a lab fixed frame, which eliminates
the difficulties associated with determination of event plane for
conventional elliptic flow analysis for non-central collisions.
It was shown in \cite{cmbhic} that a plot of values of 
$v_n^{rms}$ vs. $n$ can be 
used for directly probing various flow coefficients, in particular,  
the elliptic flow for non-central collisions \cite{flow0}, without any 
need for the determination of event-plane. 

 In this paper we continue to explore this fertile connection between 
CMBR physics and RHICE. It was shown in ref. \cite{cmbrB} that the 
presence of background magnetic field in cosmic plasma can distort the 
acoustic peaks in CMBR. This happens due to the presence of different 
waves in the plasma, with velocities depending on the angle between 
the magnetic field and the wave vector. Presence of very strong magnetic 
fields in the plasma (of order 10$^{15}$ Tesla) during early
stages in RHICE has been explored extensively recently in connection
with the exciting possibility of observing CP violation effects
\cite{cp,bcal1,bcal2}. An important effect of the presence 
of such strong magnetic fields in the plasma will be to lead to strong 
variations in velocities of different types of waves in the plasma. 
In particular the velocity varies with the angle between the wave vector
and the direction of 
the magnetic field. It is thus obvious that this may qualitatively affect 
the development of anisotropic flow. We argue that the flow coefficients 
can be significantly affected by these effects. In particular, the 
presence of magnetic field can lead to enhancement in the elliptic flow 
coefficient $v_2$ by almost 30 \%. (Note, we use $v_2$ with the present 
definitions to denote the  elliptic flow even though we do not adopt the 
conventional usage of the eccentricity for defining the corresponding 
spatial anisotropy.) This raises the interesting possibility whether a 
larger value of $\eta/s$ can be accommodated by RHIC data when these 
effects are incorporated using full magnetohydrodynamical simulations.

 An important issue here is the time scale over which the magnetic field 
remains strong. The magnetic field arising from the valence charges of the 
initial nuclei peaks to strong values for a very short time, essentially 
the passing time of the Lorentz contracted nuclei ($\sim$ 0.2 fm for RHIC 
energies). Subsequently it rapidly decays ($\sim \tau^{-2}$) \cite{bcal1}. 
In such a situation the effect of magnetic field on flow coefficient will be 
suppressed as time scale for the development of flow is several fm. Even 
at LHC, where expected magnetic field is more than an order of magnitude 
larger than that at RHIC, the effect of this initial pulse of magnetic 
field on flow coefficients may not be very significant, though due to 
various uncertainties, a full magnetohydrodynamics simulation is needed 
to investigate these issues.
 
 However, it has been recently pointed out \cite{tuchin} that magnetic fields 
of similar magnitude, as the peak value of the external field, can arise from 
induced currents due to rapidly decreasing external field. Further, in the
quasi-static approximation, the
magnetic field satisfies a diffusion equation with the diffusion
constant equal to $1/(\sigma \mu)$ where $\mu$ is the magnetic permeability
and $\sigma$ is the electrical conductivity \cite{dfusn}. It is argued in ref. 
\cite{tuchin} that due to this the magnetic field  may survive for much 
longer time, and can lead to interesting effects. 

  We will take $\mu \sim 1$ (refs.\cite{dfusn,tuchin}). However, the value 
of electrical conductivity we take is smaller than what is taken in
ref.\cite{tuchin}. We use $\sigma \simeq 0.3 T$ (= 0.3 fm$^{-1}$ for
T $\simeq$ 200 MeV) from refs.\cite{sigma}. The time scale
$\tau$ over which the magnetic field remains essentially constant 
\cite{tuchin} over length scale $L$ is, 
$\tau \simeq L^2 \sigma/4$. For $L = 5 - 6$ fm, we get $\tau \simeq
2-3$ fm. For higher temperatures $\sigma$ will be larger increasing
the value of $\tau$. (See, also, ref. \cite{leptons} for effects of
leptons on $\sigma$, though for RHICE this may not be significant.)
$\sigma$ is also expected to increase due to the effects of magnetic
field in the plasma \cite{sigmaB}, further increasing the value of $\tau$. 
 
 It is important to note that the initial magnetic field will enter the 
medium in the longitudinal direction as the medium is only about 1 fm 
thick in that direction at the initial stage. For this the relevant quantity
is the penetration depth $\delta \sim (\mu \sigma \omega)^{-1/2}$
where $\omega$ is the angular frequency of electromagnetic wave.
Initial magnetic field, being a narrow pulse of time duration
$t \simeq 0.2$ fm (typically the width of Lorentz contracted Nuclei,
for RHIC energies), can be taken to have $\omega \simeq 30$ fm$^{-1}$.
This gives the penetration depth of order 3 fm. Thus, the picture of
magnetic field diffusing through the entire region of the plasma 
with typical length scale of several fm, and lasting with high 
initial peak values for time scales of several fm, is self consistent.
(Though, note that the time scale of the decay of the field being
of same order as the system size makes the assumption of
quasi-static field, and to that extent the assumption of ideal
magnetohydrodynamics, only marginally valid.) As significant flow 
anisotropy develops also in the  time scale of order 3-4 fm \cite{oltr}, 
it is then reasonable to assume that magnetic field can 
be taken to be almost constant for this duration of flow development.  
In conclusion, induced magnetic field,
with similar magnitude as the peak value of the initial magnetic
field pulse, decays slowly and, for time scales relevant for our
model, can be taken to be approximately constant.

   With this discussion, we continue to investigate the effects of
(an approximately constant) magnetic field on sound waves in QGP produced 
in RHICE. For the non-relativistic plasma the effect of magnetic field
can be simply described in terms of three different waves \cite{cmbrB,mhd}.
There is a fast magnetosonic wave which, for small magnetic fields, becomes
the ordinary sound wave. Its velocity is given by $c_+^2 = c_s^2 + 
v_A^2 \sin^2\theta$, where $v_A = B_0/\sqrt{4\pi\rho}$ 
is the Alfv\'en velocity and $\theta$ is the angle between the wave vector
and the magnetic field ${\bf B}_0$. $c_s$ is the sound velocity and 
$\rho$ is the plasma density. Slow 
magnetosonic wave has velocity $c_-^2 = v_A^2 \cos^2\theta$, and the 
Alfv\'en wave velocity is given by $c_A^2 = v_A^2 \cos^2\theta$. These
expressions for magnetosonic wave velocities are valid for the case
when $v_A << c_s$. It was argued in \cite{cmbrB} that fast magnetosonic
waves lead to distortion of CMBR acoustic peaks while the slow
magnetosonic waves may lead to long period modulation of the peaks.

 For the situation of plasma in RHICE one needs to consider
the situation of relativistic magnetohydrodynamics. We will assume
the applicability of magnetohydrodynamical description for the plasma
at RHICE and restrict our consideration to wavelengths much larger
than the Debye screening length. For the relativistic case, the
expressions for the velocities of these  waves are given below
(we use natural units with $c$ = 1). For the following discussion, 
we have followed ref.\cite{mhd}. \\

 {\bf Alfv\'en waves}

 Phase velocity ${\bf v}_{ph}$ and group velocity ${\bf v}_{gr}$
of these waves are 

\begin{equation}
{\bf v}_{ph} = {B_0 \cos\theta \over \sqrt{\omega_0}} {\bf n}, 
\qquad {\bf v}_{gr} = {{\bf B_0} \over \sqrt{\omega_0}} ,
\end{equation}  
  
where ${\bf n} = {\bf k}/k$, ${\bf k}$ is the wave vector and
$\theta$ is the angle between ${\bf n}$ and the magnetic field
${\bf B_0}$. $\omega_0 = \rho_0 h_0 + B_0^2$ where $h$ is the specific
enthalpy (subscript denoting the background values) defined as \cite{mhd}
$\rho h \equiv \rho c^2 + \rho \epsilon + P$. Here, $\rho \epsilon$ is the 
internal energy, $\rho c^2$ is the rest mass energy and $P$ is the pressure. 
For ultra-relativistic case we take $\rho h = 4 P$. \\

{\bf Magnetosonic waves}

These are the waves which are relevant for our case of discussion of 
flow as they involve density perturbations. Phase velocities for these
waves are given by 

\begin{equation}
{\bf v}_{ph} = v_{ph} {\bf n} = {\bf n} ({1 \over 2}[(\rho_0 h_0 /\omega_0)
c_s^2 + v_A^2])^{1/2} (1 + \delta \cos^2\theta \pm a)^{1/2} .
\end{equation}

Here $+$ and $-$ signs correspond to the fast and slow magnetosonic
waves respectively, $v_A = B_0/\sqrt{\omega_0}$ is the Alfv\'en speed, 
and $\delta$ and $a$ are defined below.

\begin{equation}
a^2 = (1 + \delta \cos^2\theta)^2 - \sigma \cos^2\theta , \\
\end{equation}
\begin{equation}
\delta = {c_s^2 v_A^2 \over [(\rho_0 h_0/\omega_0)c_s^2 + v_A^2]} , 
~~ \sigma = {4c_s^2 v_A^2 \over [(\rho_0 h_0/\omega_0)c_s^2 + v_A^2]^2} .\\
\end{equation}
 
($\sigma$ here should not be confused with the conductivity discussed
above.) For propagation of density perturbations, as relevant for the 
evolution of flow anisotropies, the relevant wave velocity is the group 
velocity for the magnetosonic waves,

\begin{equation}
{\bf v}_{gr} = v_{ph}\left[{\bf n} \pm {\bf t} {[\sigma \mp 2\delta (a
\pm (1 + \delta \cos^2\theta))]\sin\theta \cos\theta \over 2(1 +
\delta \cos^2\theta \pm a)a}\right] .
\end{equation}

Here ${\bf t} = [({\bf B_0}/B_0) \times {\bf n}] \times {\bf n}$, and
again the upper and lower signs ($\pm$ or $\mp$) correspond to the
fast and the slow magnetosonic waves respectively. For a given magnetic 
field ${\bf B_0}$, the direction of ${\bf n}$ can be varied to generate 
group velocities of these waves in different directions. Fig.1 shows
a typical situation of various vectors in Eq.(5) expected in RHICE.
It is important to note that the direction of ${\bf v}_{gr}$ depends
on the relative factors multiplying ${\bf n}$ and ${\bf t}$ in Eq.(5).
This in turn depends on properties of the plasma like energy density.
Thus due to the presence of spatial gradients in RHICE, even along a 
fixed azimuthal direction, we will expect the direction of 
${\bf v}_{gr}$ to keep varying with the radial distance. This can lead 
to the development of very complex flow patterns, possibly leading to 
generation of vorticity.

\begin{figure}
\leavevmode
\epsfysize=6truecm \vbox{\epsfbox{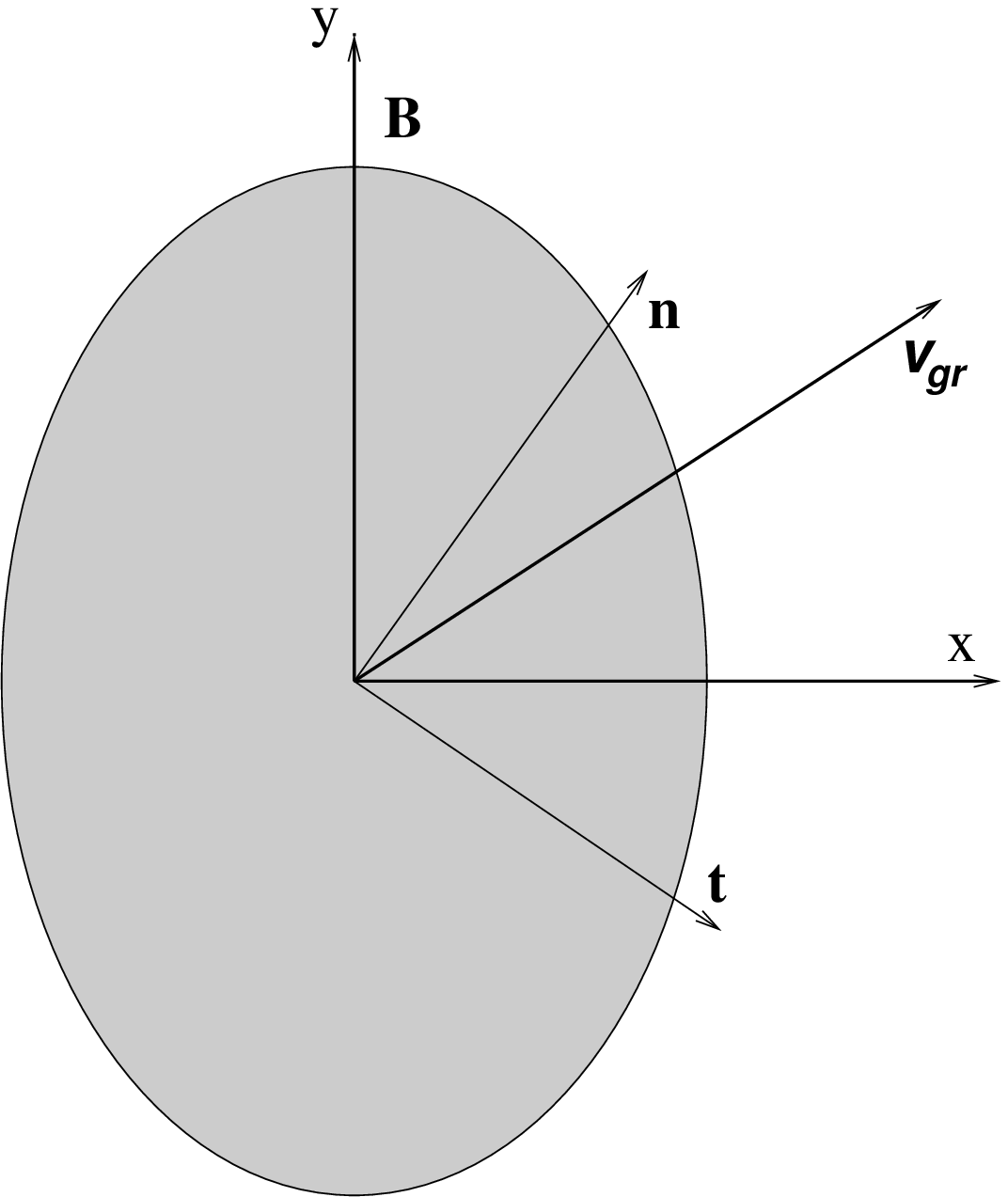}}
\caption{}{Typical situation expected in RHICE. Magnetic field
points in y direction and the direction of the group velocity 
${\bf v}_{gr}$ is obtained from ${\bf n}$ and ${\bf t}$ via Eq.(5).}
\label{Fig.1}
\end{figure}

  For the strength of the magnetic field expected in RHICE, we use
the estimates given in refs.\cite{bcal1,bcal2} for Au-Au collisions
at RHIC. For first estimates, we use a simple parametrization and take 
the magnetic field to be proportional to the impact parameter
(with somewhat larger magnitudes than in ref. \cite{bcal1,bcal2} to 
illustrate the effects).

\begin{equation}
B_0 = 10^5 {b(fm) \over 10} {\rm MeV^2} . 
\end{equation}

 This corresponds to the maximum magnetic field of about $10^{15}$ 
Tesla and the scale of 10 fm is taken from \cite{bcal2} (where
time integrated value of $eB_0$ is calculated, as coupling to charge
particles is of interest there) as the approximate limiting value of the 
impact parameter $b$ up to which proportionality with $b$ is expected. 
The effect of magnetic field on wave propagation here comes from an effective 
magnetic pressure arising from the freezing of magnetic field lines in 
the plasma in the magnetohydrodynamic limit. Note that the importance of 
charges (quarks) here is only in establishing the magnetohydrodynamic 
conditions such that the electric field in the comoving frame vanishes.
There is no direct pressure generated by interaction of magnetic field
with quarks. (Thus it is of no relevance that gluons, which do not interact
with the magnetic field, dominate the energy density of the plasma.) The 
distortions of magnetic field lines in the presence of density perturbations
cost energy leading to an extra contribution to pressure from
the presence of magnetic field. This is what is responsible for 
increasing the effective sound speed as given above. 

   We will assume a magnetic field with magnitude $B_0$ given 
above which is uniform over the region of the plasma, and as we have
discussed above, is approximately constant for the early stages of
time scale of few fm. This early time duration is important for the 
evolution of flow anisotropies \cite{oltr}. The magnetic field
has direction in the transverse plane, 
normal to the direction of the impact parameter. From the expression of
${\bf v}_{gr}$ in Eq.(5) we see that an important factor is the ratio
$B_0^2/P$ where $P$ is the pressure of the plasma which we take to be 
the quark-gluon plasma with two light flavors with the pressure
given by $P = {37 \over 90}\pi^2 T^4$. Again, to illustrate the effects
of the magnetic field on flow, we consider the situation at a lower
value of the temperature $T = 180$ MeV. As the important effects occur
for strong magnetic field which occurs for large impact parameter, a 
lower value of $T$ (compared to what is expected in central collisions)
may not be very unreasonable.  We are considering the effects
of magnetic field relevant for Au-Au collisions at RHIC on somewhat
larger side of the estimates. However, for higher energy collisions,
e.g. at LHC, larger values of $B_0$ (by an order of magnitude
compared to RHIC) should be routinely expected, see e.g. \cite{lhc}.
  
  We follow the procedure described in \cite{cmbhic} for
calculating $v_n^{rms}$ using HIJING \cite{hijingp}. We start with the 
initial transverse energy density $\epsilon_{tr}$ distribution for Au-Au 
collision at 200 GeV/A center of mass energy from HIJING. For details, see
ref.\cite{cmbhic}.
We assume that the hydrodynamic description becomes applicable by 
$\tau =\tau_{eq}$, which we take to be 1 fm and calculate the 
anisotropies in the fluctuations in the spatial extent $R(\phi)$ at 
this stage, where $R(\phi)$ represents $\epsilon_{tr}$ 
weighted average of the transverse radial coordinate in the angular bin 
at azimuthal coordinate $\phi$. As emphasized above, 
angle $\phi$ is taken in a lab fixed coordinate frame. 
We divide the region in 50 - 100 bins of azimuthal angle $\phi$, and 
calculate the Fourier coefficients of the anisotropies in ${\delta R}/R 
\equiv  ({\bar R}- R(\phi))/{\bar R}$ where $\bar R$ is the angular 
average of $R(\phi)$.  Note that in this way we are representing
all fluctuations essentially in terms of fluctuations in the boundary of
the initial region.  We use $F_n$ to denote Fourier coefficients for 
these spatial anisotropies, and use $v_n$ to denote $n_{th}$ 
Fourier coefficient of expected momentum anisotropy in ${\delta p}/p$
defined in the lab frame.  We have generated events using HIJING and we 
present sample results for Au-Au collision at 200 GeV/A center of mass 
energy. In all the plots, the averages are taken over 1000 events.

 In \cite{cmbhic}, the root mean square values $v_n^{rms}$ 
of the flow Fourier coefficients were obtained from spatial 
$F_n$s simply by using proportionality factor of 0.2 (with a minus
sign as $F_n$ will be negative).
 We include the effect of magnetic field and the resulting
angle dependent velocity of the (fast) magnetosonic wave by replacing this
proportionality factor to 0.346 $\times {\bf v}_{gr}$. Here the group
velocity of the (fast) magnetosonic wave ${\bf v}_{gr}$ (Eq.(5)) changes with
the angle in the event plane. The assumption here is that the flow
coefficients are proportional to the sound velocity \cite{oltr}. (Ignoring
that now the time scale of the development of flow may also vary with 
the azimuthal angle). The factor of 0.346 is chosen so that
the proportionality constant becomes 0.2 for zero magnetic
field case with the usual sound velocity $c_s = 1/\sqrt{3}$.

 In Fig.2 we show the plots of $v_n^{rms}$ for  different values
of impact parameter $b$ with magnetic field given by Eq.(6)
(Solid curves) and without magnetic field (dashed curves). These plots
show that magnetic field can strongly affect values of $v_n^{rms}$.

\begin{figure}
\begin{center}
\leavevmode
\epsfysize=8truecm \vbox{\epsfbox{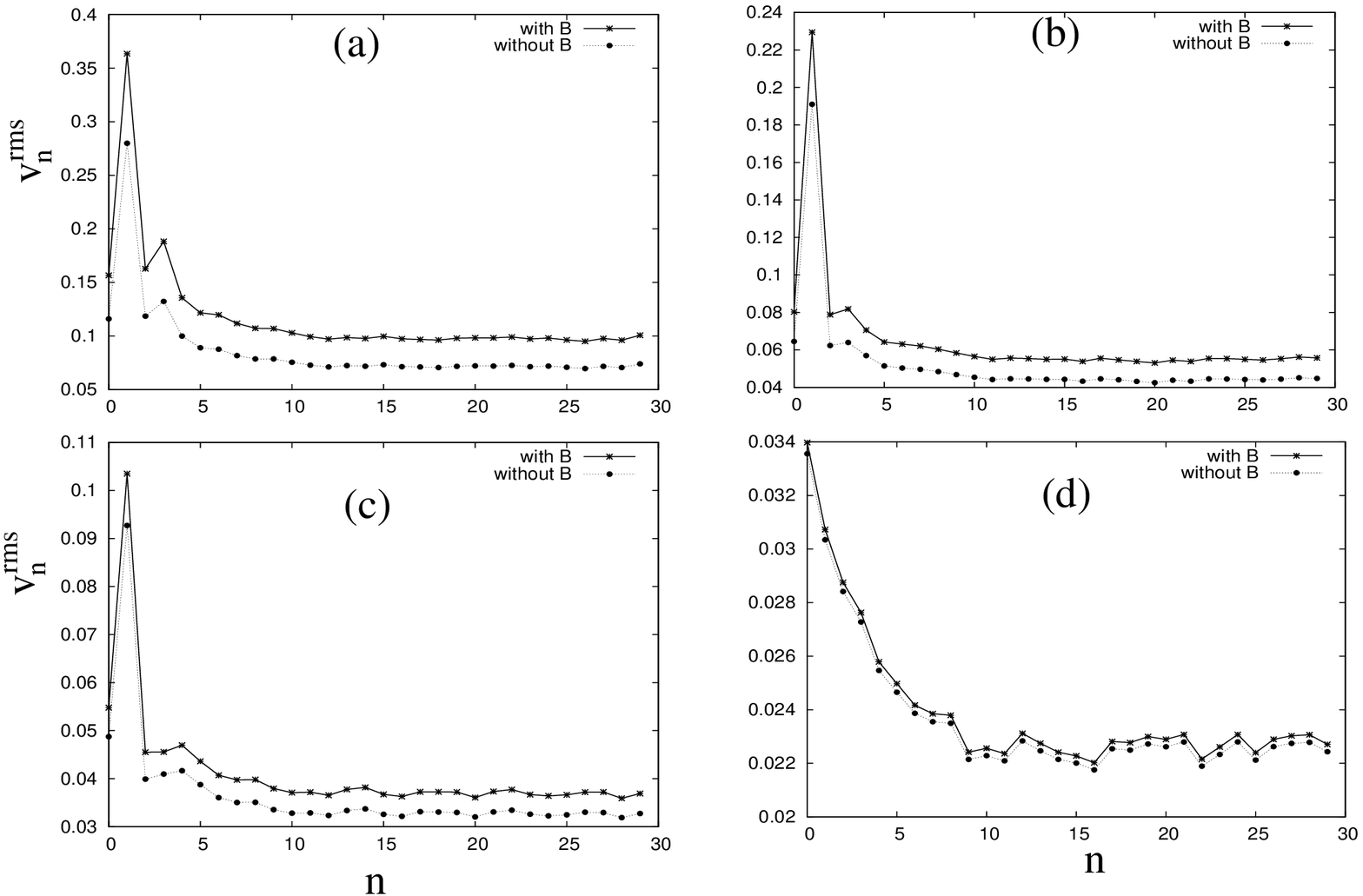}}
\end{center}
\vskip -0.2in
\caption{}{Plots of $v_n^{rms}$ for different values of impact
parameters $b$. Solid (dashed) curves show the plots in the
presence (absence) of magnetic field. (a),(b),(c),(d) correspond to
the values of $b$ = 10,8,6,2 fm respectively.} 
\label{Fig.2}
\end{figure}

  We have also calculated average values of flow coefficients in
the event plane.  Fig.3 shows the plots of the
average flow coefficients $v_n$ for $b$ = 10 fm. Solid curve shows 
the plot with the presence of magnetic field and the dashed curve
shows the plot in its absence. Note that both the curves approach almost
zero value beyond $n \simeq $ 10 (as higher $v_n$s lose correlation with 
the event plane).

\begin{figure}
\begin{center}
\leavevmode
\epsfysize=5truecm \vbox{\epsfbox{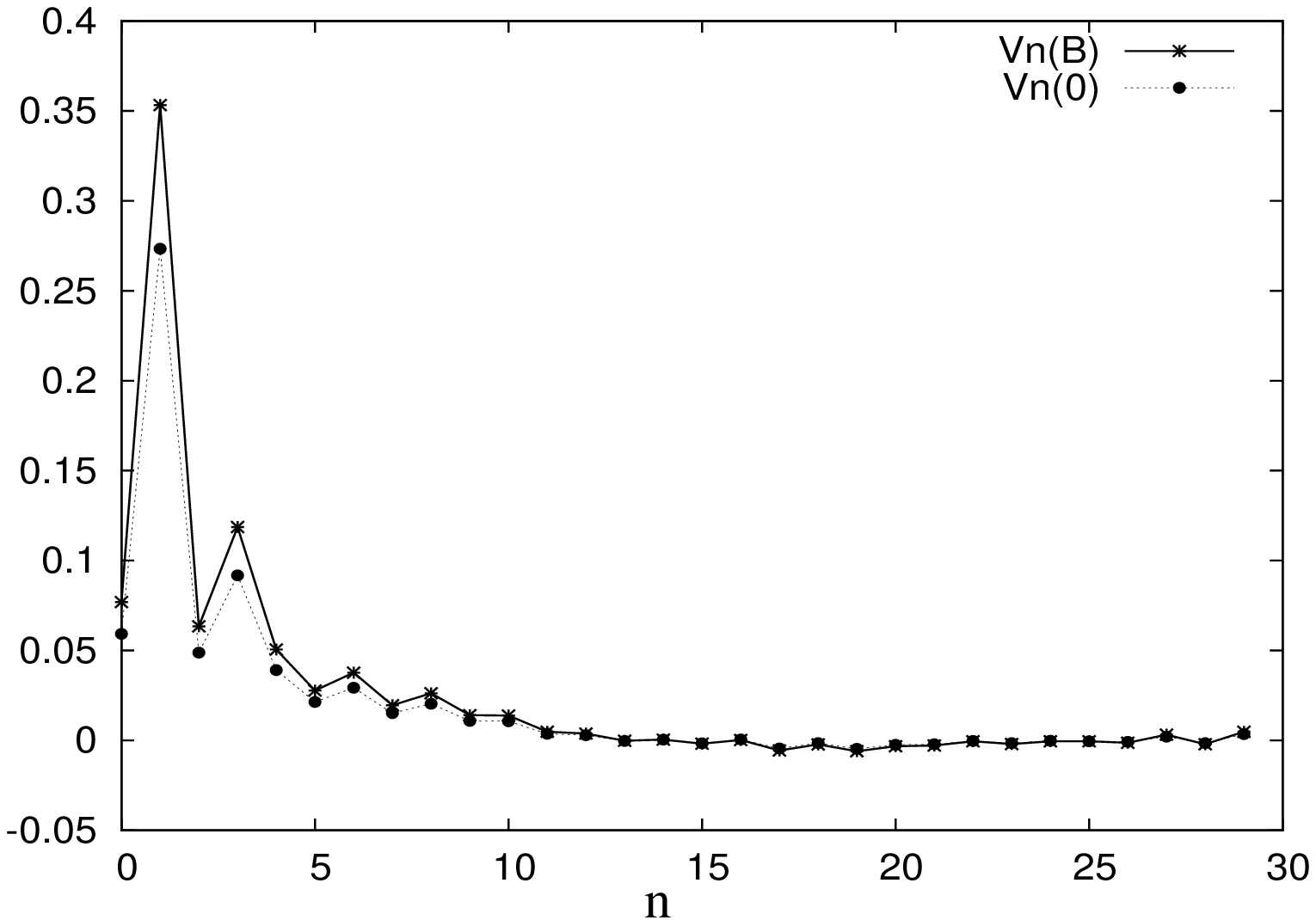}}
\end{center}
\vskip -0.2in
\caption{}{Plots of the average flow coefficients $v_n$ for $b$ = 10 fm. 
Solid curve shows the plot with the presence of magnetic field and the 
dashed curve shows the plot in its absence.}
\label{Fig.3}
\end{figure}

It is important to study the effect of magnetic field on the
elliptic flow $v_2$. In Fig.4 we show the plot of the ratio of
$v_2$ with and without the magnetic field, i.e. $v_2(B)/v_2(0)$
as a function of the impact parameter $b$. Elliptic flow itself
varies with $b$ and this ratio helps us in separating the effect
of the magnetic field on $v_2$ as the field varies with $b$.
Note that this ratio becomes as large as 1.3 for $b$ = 10 fm
(for which the magnetic field takes its largest value in Eq.(6)).
This strong enhancement in the value of $v_2$ is important especially
as it raises the interesting possibility whether a larger value
of $\eta/s$ can be accommodated by RHIC data when proper
accounts of magnetohydrodynamics are incorporated in flow
calculations. 

\begin{figure}
\begin{center}
\leavevmode
\epsfysize=5truecm \vbox{\epsfbox{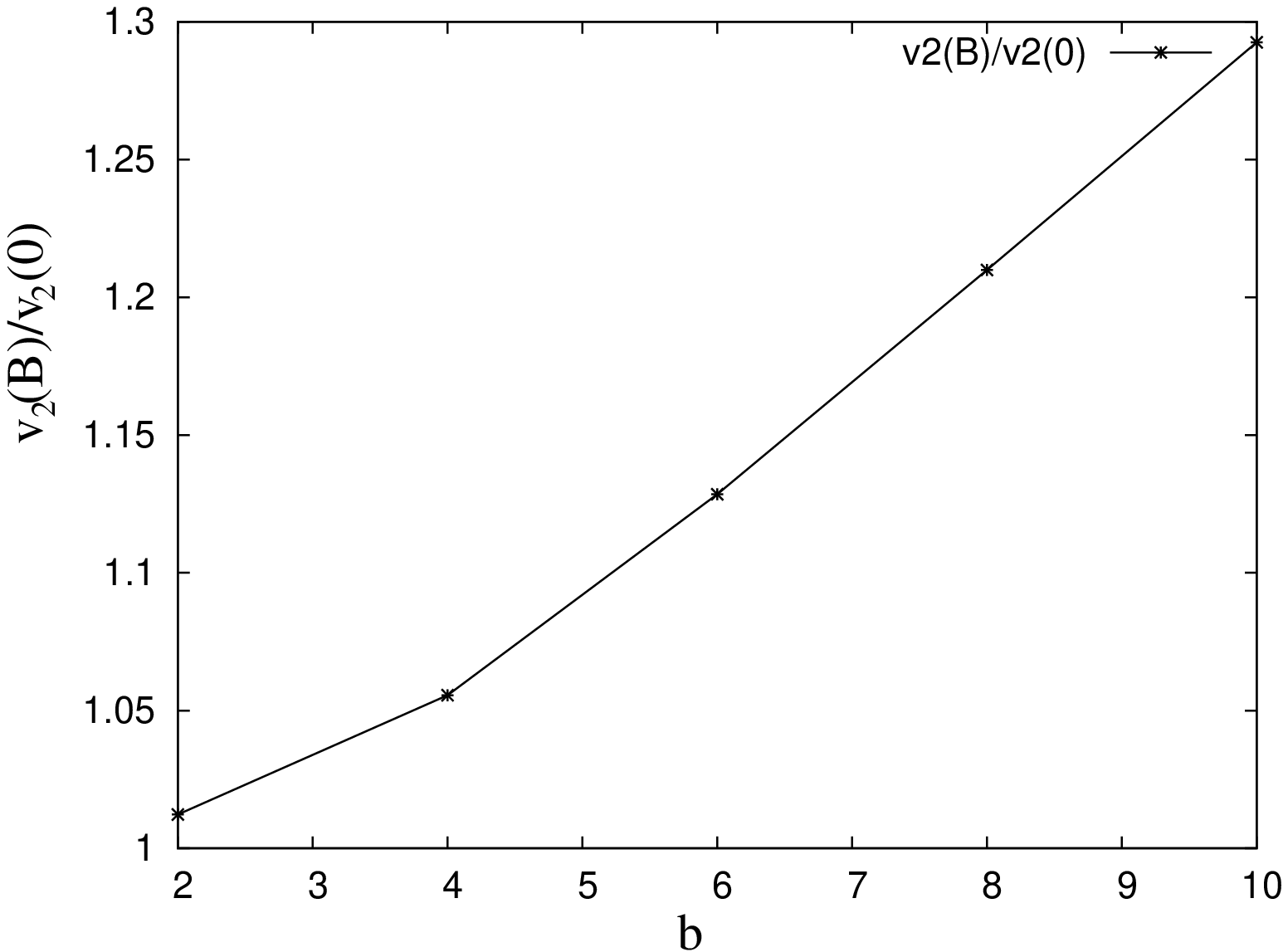}}
\end{center}
\vskip -0.2in
\caption{}{Plot of the ratio of $v_2$ with and without the magnetic field, 
as a function of the impact parameter $b$ (in fm).}
\label{Fig.4}
\end{figure}

 Due to anisotropy of magnetosonic wave velocity, the radial flow
itself will get modified. To incorporate this effect we modulated
the initial spatial profile with a suitable weight factor proportional
to ${\bf v}_{gr}$. However, its effects were insignificant. It still
remains a possibility that for much stronger magnetic fields
non-trivial flow anisotropies may arise even with almost isotropic 
initial conditions (though, for central collisions one expects very small
magnetic fields). Also, note that the direction of ${\bf v}_{gr}$
depends on the relative weights of factors multiplying ${\bf n}$
and ${\bf t}$ in Eq.(5) which depends on quantities like the plasma
energy density etc. Thus due to the presence of spatial gradients
of plasma density in RHICE, even along a given azimuthal direction 
in the transverse plane the direction of ${\bf v}_{gr}$ will
keep changing. In particular, fluctuations of energy density
will lead to fluctuations in ${\bf v}_{gr}$ as well. 
Clearly due to such phenomena one expects a complex
pattern of flow, even possibly leading to vorticity, to develop in 
RHICE than just radial flow and flow
anisotropies. We have also checked the effects of magnetic field on  other 
features of the plots of $v_n^{rms}$ discussed in \cite{cmbhic}, in 
particular on the acoustic peaks. The effects are similar to what is 
shown in Fig.2 hence we do not show it here. It remains to be explored
how the slow magnetosonic waves affect various features of the
plots of $v_n^{rms}$ in view of its proposed long period modulation of
the acoustic peaks in CMBR. The role of Alfv\'en waves also needs to
be explored in the evolution of various fluctuations in RHICE 
(especially in view of their effects on CMBR \cite{cmbalfven}).
Our approach in this work has been
to present rough estimates of various possible effects of the
presence of the magnetic field in RHICE. Detailed magnetohydrodynamical
simulations are needed to probe these effects. Especially exciting
will be the possibility of large effects at LHC energies where 
large magnetic fields (by almost an order of magnitude compared to RHIC)
are expected \cite{lhc}.

  We are very grateful to Abhishek Atreya, Anjishnu Sarkar, Uma Shankar
Gupta, and Trilochan Bagarti for useful discussions.


\end{document}